# $O_2^-$ Ion Mobility in Dense Ne Gas: the Free Volume Model

A. F. Borghesani
Department of Physics & Astronomy and CNISM
University of Padua
Padua, Italy

F. Aitken
G2ELab, CNRS, Grenoble INP
University Grenoble Alpes
Grenoble, France

## ABSTRACT

We report data of the $O_2^-$ ion mobility in neon gas over broad density and temperature ranges along with its theoretical description in terms of the thermodynamic, free volume model that has successfully been adopted for the interpretation of electron and ion mobility in superfluid and normal helium. The free volume model, which is aimed at computing the free volume accessible for the ion motion, along with the Millikan-Cunningham slip factor correction, is able to describe the ion mobility in the crossover region connecting the dilute gas regime described by the classical Kinetic Theory to the high density region ruled by the laws of hydrodynamic transport.

## 1   INTRODUCTION

**THE** drift motion of ions induced by an electric field in dense gases and liquids is very important to understand and design processes in many applicative fields, such as those, e.g., involving low-temperature plasmas, as well as investigating the microscopic interaction mechanisms between charges and neutrals for basic research. Ion transport takes places in several natural and industrial processes as, for instance, in biology, chemical synthesis, electrical discharges, and is also one of the fundamental mechanisms underlying the operations of high-energy physics detectors. The understanding of the physical chemistry of the atmosphere depends on the knowledge of the fundamental mechanisms of the interaction of the ions with the particles of the neutral species [1-3].

The experimental data on the ion mobility $\mu$ obtained in dilute gases can be used to extract information on the ion-neutral interaction potential within the framework of classical Kinetic Theory or $\mu$ can be predicted from the knowledge of the potential. For thermal ions, i.e., when the electric field is so weak as not to significantly heat them, the scattering cross section is given as $\sigma = 4\pi R^2$, in which $R$ is the hard-sphere radius of the ion-neutral interaction, and the density-normalized- or reduced mobility is expressed as

$$\mu N = \frac{3}{2\sigma}\left(\frac{\pi}{2m_r k_B T}\right)^{\frac{1}{2}}, \qquad (1)$$

in which $N$ is the gas number density, $m_r$ is the ion-neutral reduced mass, $k_B$ is the Boltzmann constant, and $T$ is the absolute temperature [3].

In liquids, owing to their large density and to the very small mean free path of neutrals with respect to the ion size, the mobility of singly charged ions is described within the frameworks of Hydrodynamics and is given by the Stokes' formula

$$\mu = \frac{e}{6\pi\eta R}, \qquad (2)$$

in which $e$ is the ion charge, $\eta$ is the medium dynamic viscosity, and $R$ is the ion effective hydrodynamic radius [3]. The investigation of the ion mobility in this regime in media such as normal and superfluid helium (for a review, see Ref. [4]), liquid hydrocarbons [5], and cryogenic liquids [6-8], helped elucidating the microscopic structure and dynamics of liquids.

Unfortunately, a theoretical description of the mobility in the intermediate density range between the two asymptotic transport regimes is missing though this range can experimentally be explored by investigating ions in supercritical gases. Actually, especially close to the critical temperature, the gas density can be varied quite at will by changing pressure. In this way, the Knudsen number $K_n = \ell/R$, i.e., the ratio of the mean free path $\ell$ of the neutrals to the ion size, ranges from low values, typical of the hydrodynamic transport regime, to the very large ones featured by the kinetic regime.

In particular, the mobility of negative ions, in spite of their relevance in many topics, is not so extensively studied as that of positive ions because their production is not easy. Anions are formed via the resonant electron attachment process to electronegative impurities, such as $O_2$. The collision of an excess electron with such a molecular impurity leads to the formation of a temporary negative ion in a vibrationally excited state that quickly decays by autodetachment. Only a fraction of ions is stabilized by collisions with a third body carrying away the excess energy, thereby giving birth to long-lived anions whose mobility can be measured [9,10]. Actually, measurements of the resonant attachment of thermal or epithermal electrons to $O_2$ impurities have been carried out in He and Ne and the vibrational structure of the ion has been resolved [11-14]. This mechanism is quite inefficient and the amount of ions that can be produced is fairly small. In any case, the formation of stable anions strongly depends on the environment as a consequence of the delicate balance between the long-range polarization interaction and the short-range repulsive exchange forces. This balance gives origin to a state that cannot be adiabatically obtained as a result of a simple addition of the ion to the host medium, and the resulting ion structure is more complicated than that of cations.

The most important attaching species, both because of its abundance and because of its relevance to the life on Earth for its role in the physical chemistry of the atmosphere, is $O_2$. Its overall availability as impurity makes possible to study its ion $O_2^-$ in many gases, thereby detecting specific features of the host gas. Only few experiments have been aimed at measuring the $O_2^-$ ion mobility in liquid Ar, Kr, and Xe [6], and in supercritical dense Ne, Ar, and He [13,15-17] as a function of density and temperature. Measurements in supercritical gases were carried out in the crossover region between the two limiting transport regimes and lead to the conclusion that neither the Kinetic Theory nor the Hydrodynamic Theory are able to explain the experimental outcome.

Recently, a thermodynamic model has been developed to successfully describe the mobility of electron bubbles and cations in superfluid and normal liquid He. The model is based on the free volume concept and provides a thermodynamic equation of state to compute the temperature and density dependence of the volume available to the ion motion. The size of the free volume is assumed to be the effective ion hydrodynamic radius and the mobility is computed via the Stokes' equation. The prediction of the free volume model can be extended to the crossover and dilute gas regions by using a modified form of the Millikan-Cunningham slip correction factor [18-21].

We successfully adopted the free volume model to describe the mobility of oxygen ions in dense supercritical Ne gas [22]. We now present here new measurements in dense supercritical Ne covering a temperature range from just above the critical one to well above room temperature in order to complete the assessment of the validity of the free volume description of the crossover region between the kinetic and the hydrodynamic transport regimes.

## 2 EXPERIMENTAL RESULTS

The experiment is based on the well-known pulsed Townsend photo-injection technique and has technically been detailed elsewhere [15]. We only recall its main features. The experimental cell is designed to contain gas up to 10 MPa and can be cooled down to 25 K. The thermoregulation is within 0.01 K. The Ne gas contains $O_2$ impurities in a concentration of a few tens of parts per million. The ions are produced within the first few Å of the drift space by resonant electron attachment and drift under the action of a uniform electric field applied to two plain parallel electrodes separated by a distance of $10^{-2}$ m. The ion current is integrated to maximize the signal-to-noise ratio.

We investigated the temperature range 45 K<$T$<330 K. On the isotherm $T$=45 K, closest to the critical temperature $T_c$=44.38 K, we reached a density as high as $N$=230·$10^{26}$ m$^{-3}$, well above the critical density $N_c$=144.4·$10^{26}$ m$^{-3}$. At all higher isotherms, the density was always as high as to fall within the crossover region.

The reduced drift electric field $E/N$ never exceeds a few tens of mTd (1 Td = $10^{-21}$ Vm$^2$) so that ions always are in thermal equilibrium with the gas and $\mu$ does not depend on the field $E$ even at the lowest density, as shown in Figure 1.

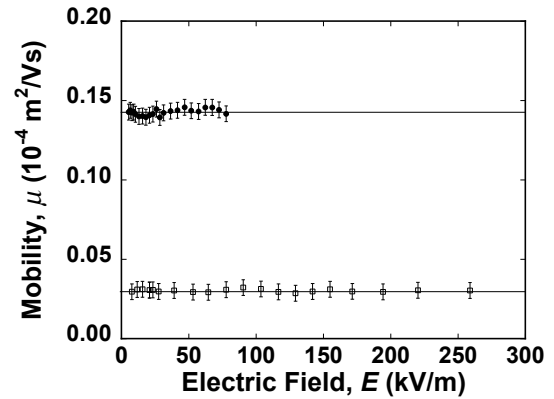

**Figure 1.** $\mu$ vs $E$ in Ne at $T$=64.4 K for $N$=5.5·$10^{26}$ m$^{-3}$ (closed symbols) and $N$=27.3·$10^{26}$ m$^{-3}$ (open symbols).

Although being field independent, $\mu$ does not conform to the prediction of the Kinetic Theory, Equation (1), because the reduced mobility $\mu N$ shows a weak, almost linear density dependence, as shown in Figure 2. This behavior is common to other noble gases such as Ar and He [16,17].

The inadequacy of both the predictions of Kinetic Theory and of Hydrodynamics is best highlighted by the experimental results obtained on the 45 K isotherm, shown in Figure 3, for which the highest densities are reached. We note that the Kinetic Theory predicts $\mu N$ independent of $N$ at constant $T$, a behavior not shown by the experimental data.

Also Hydrodynamics fails at describing the ion mobility, even when $N$ is so large that $K_n \ll 1$. Actually, in presence of the strong perturbation exerted by the ion on the surrounding medium, the Stokes' hydrodynamic formula must be modified so as to yield

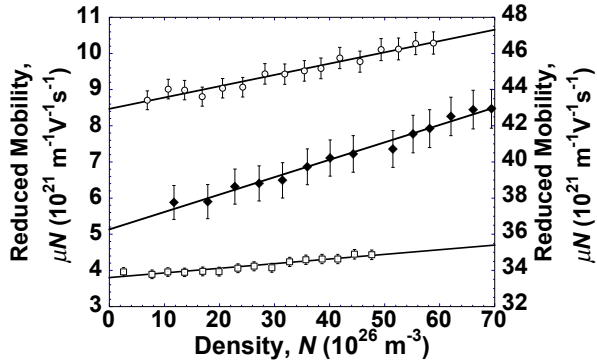

**Figure 2.** $\mu N$ vs $N$ in Ne at $T$=95.9 K (circles) [16] and Ar at $T$=180 K (squares) [17] (left scale). $\mu N$ vs $N$ in He at $T$=77.6 K (diamonds) [16] (right scale). The lines are only to guide the eyes.

$$\mu = \frac{e}{6\pi\eta FR}. \qquad (3)$$

The correction factor $F$ is obtained by integrating the Navier-Stokes equations in presence of the strong density and viscosities nonuniformities around ions induced by electrostriction [15]. The dash-dotted curve in Figure 3 is obtained with $R$=3.25 Å and shows that the Stokes' formula exploited to describe the data. The position of the density profile maximum is chosen as the ion hydrodynamic radius and the viscosity is evaluated at the density of the maximum. The results are shown in Figure 3 as the dotted line. The agreement with the data is reasonably good for $N > 150 \cdot 10^{26}$ m$^{-3}$ but the computations fail at reproducing the data for smaller densities. It appears that even taking into account the microscopic structure of the ion by including both polarization and exchange interactions it is not possible to describe the ion mobility over the whole crossover region.

Even recent Molecular Dynamics calculations [25] (open circles in Figure 3) are relatively good at describing the mobility in the high density region but fail to describe both the crossover as well as the low density behavior of the data.

An additional confirmation of the limitations of Kinetic Theory is its failure at describing the zero-density limit of the reduced mobility $(\mu N)_0$, given by the Langevin formula

$$(\mu N)_0 = 1.8 \frac{\epsilon_0}{\sqrt{m_r \alpha_s}}, \qquad (4)$$

in which $\epsilon_0$ is the vacuum permittivity and $\alpha_s = 4.4 \cdot 10^{-41}$ Fm$^2$ is the Ne static polarizability. The Langevin prediction is a factor 2 larger than the experimental data, as shown in Figure 4. Moreover, the experimental data depend on temperature in a way which is not predicted by Equation 4.

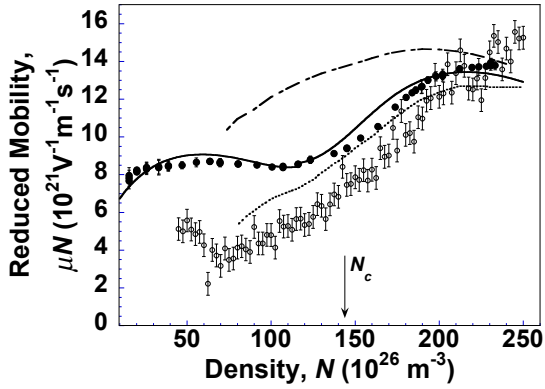

**Figure 3.** $\mu N$ vs $N$ in Ne at $T$=45 K. Closed circles: experiment. Dash-dotted line: hydrodynamic mobility in a nonuniform fluid [15]. Dotted line: ionic bubble model [22]. Open circles: Molecular Dynamics simulations [25]. Solid line: present free volume model [22].

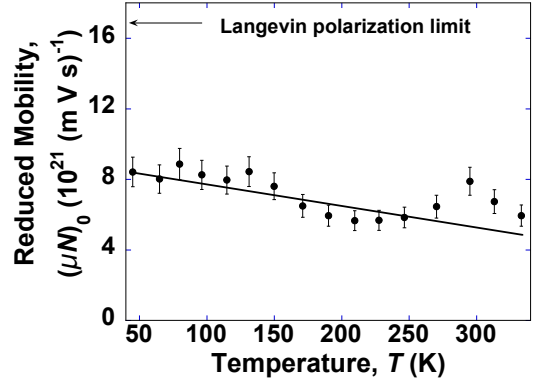

**Figure 4.** $(\mu N)_0$ vs $T$ in Ne. The solid line is computed by assuming that the slip correction factor temperature dependence is given by $A(T)$ (see text).

with the effective hydrodynamic radius $FR$ only agrees roughly with the data at the highest densities.

A better rationalization of the 45 K data is achieved by accounting for the structure of the anion [23]. In addition to the solvation shell originated by electrostriction [24] the excess electron in the ion creates a void around it because of the short-range repulsive exchange forces. The ion transport is determined by this empty void-solvation shell structure that appears to be independent of the ionic species. The void radius and the gas density profile established around the ion are computed by minimizing the free energy of the complex. In Ne close to $T_c$ size and density of the solvation shell cannot be neglected, $K_n$ is very low and the hydrodynamic formula is

## 3   THE FREE VOLUME MODEL

Owing to the shortcomings of Kinetic Theory and Hydrodynamics for the description of the mobility, a heuristic model has recently been developed that rationalizes the electron and cation mobility in normal liquid, superfluid, and supercritical He, and can also be used for negative ions [18-21]. It aims at computing the equation of state (EOS) of the free volume $V_f$ available for the ionic motion through the host medium. The ions are considered to be a solute dissolved in the medium that is considered to act as a solvent, The free volume is given by $V_f = V - b$, where $V$ is the macroscopic volume occupied by the gas and $b$ is the covolume of the atoms of the solvent. The covolume of the ions is negligible because of their extremely low concentration.

The model issues a thermodynamic description of the free volume per particle via a van der Waals-like EOS

$$V_s = \frac{V_f}{N} = \frac{k_B T}{P + \Pi}. \qquad (5)$$

$P(N,T)$ is the hydrostatic pressure and $\Pi$ is the excess internal pressure term which accounts for the excess attractive potential energy contributions in the systems. The attractive interactions in the pure fluid are already accounted for by the ordinary pressure. The linear size of this free volume per particle is related to the effective hydrodynamic radius $R$ of the ions in the Stokes' formula. The analytic relationship between $R$ and $V_s$ has to be determined by enforcing the agreement of the theoretical prediction with the experimental data because the size of the solvation shell surrounding the ions depends on the system compressibility.

In order to extend the predictive ability of the model to the crossover and low-density regions, characterized by large Knudsen number values, in which most of the experimental data have been measured, to the high-density region, in which the hydrodynamic description is valid because $K_n \ll 1$, the Stokes' formula is modified by introducing the empirical Millikan-Cunningham slip factor correction $\phi[K_n(N,T)]$ so as to yield [3]

$$\mu N = \frac{eN}{6\pi\eta R}[1 + \phi[K_n(N,T)]]. \qquad (6)$$

The slip correction factor must vanish in the high density limit and, for large Knudsen number, must behave as to yield Equation (1).

The extremely broad density range spanned by the data on the 45 K isotherm ($0.17 \leq N/N_c \leq 1.7$), shown in Figure 3, allows us to fix the values of all the model parameters. The reduced mobility shows a shallow minimum at a density $N \approx 115 \cdot 10^{26}$ m$^{-3}$ caused by a maximum of the hydrodynamic radius. It is reasonable that the hydrodynamic radius size is related to the medium compressibility. If we assume $\Pi = \alpha N^2$ with $\alpha = 0.8937$ MPa·nm$^6$ the system compressibility is maximum at the required density. It is worth noticing that the same value of α yields the best agreement between theory and experiment also in liquid and supercritical He [18]. It thus appears that α, which takes into account all the excess attractive interactions in the system, is system independent and universal. We infer that this universality might be a manifestation of the Law of Corresponding States that follows from the use of a van der Waals-like EOS.

We expect that the hydrodynamic radius $R$ is a function of the scaled free volume per particle at constant temperature $T$. The analytic form for $R$ must guarantee that $R$ is maximum at the abovementioned density. We have chosen the following form

$$\frac{R - R_0}{R_0} = \frac{(V_0/V_s)^{\epsilon_1}}{1 + \gamma(V_0/V_s)^{\epsilon_1}\exp[-K(T)(V_0/V_s)^{\epsilon_2}]} \qquad (7)$$

Here, $R_0$ is the hard-sphere radius of the ion-host atom interaction and $V_0$ is the free volume per particle of a suitable reference state. The scaled free volume per particle is given by

$$V_0/V_s = \delta \frac{P + \Pi}{T} \qquad (8)$$

$K(T)$ is a function that roughly decreases in an exponential way with temperature, as shown in Figure 5. Its behavior guarantees that $R \to R_0$ at high $T$ and low $N$. The slip factor correction is a function of the Knudsen number $K_n = \ell/R$. The mean free path $\ell$ is obtained from the viscosity $\eta$ by using

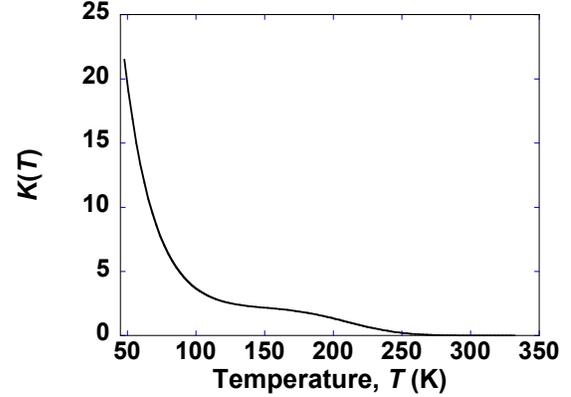

**Figure 5.** Temperature dependence of the function $K(T)$.

its Kinetic Theory expression as

$$\ell = \frac{3\eta}{m_r \bar{v} N}, \qquad (9)$$

in which $\bar{v} = (8k_B T/\pi m_r)^{1/2}$ is the mean thermal velocity. As an example, the mean free path is shown in Figure 6 as a function of the density for $T = 209$K.

As the mean free path is very large compared to $R_0$ except for the highest densities, i.e., for $K_n \gg 1$, we propose that the slip correction factor is simply given by

$$\phi(N,T) = A(T)\frac{N_c}{N}. \qquad (10)$$

in which $A(T)$ is a weakly temperature dependent function. It is determined in such a way that the free volume model reproduces the temperature dependence of the zero-field reduced mobility reported in Figure 4.

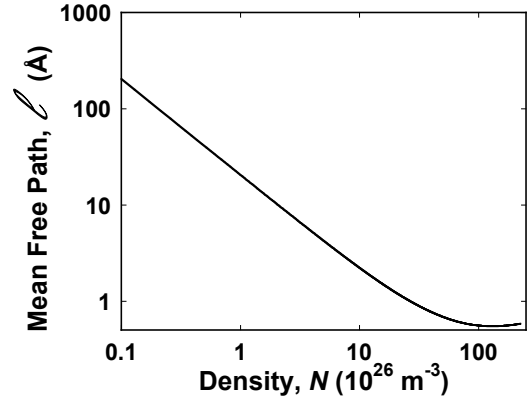

**Figure 6.** Density dependence of the mean free path at $T$=209 K.

By enforcing agreement with the experimental data for the 45 K isotherm, the adjustable parameters are set to the values: $R_0 = 3.15$Å, $\epsilon_1 = 1$, $\epsilon_2 = -1$, $\gamma = 6.3$, and $\delta = 0.15$. We once more note that the parameter $\delta$ has the same value required to describe the He case and appears to be system independent. Once more, we believe that this apparent universality is due to the van der Waals-like description of the EOS. The prediction of the free volume is shown as the thick solid line in Figure 3. We note the good agreement with the experimental data over the whole density range.

It thus appears that the free volume model is able to bridge the kinetic- and hydrodynamic transport regime by computing a correct density dependence of the effective hydrodynamic radius

$$R_{\text{eff}} = \frac{R}{1 + \phi(N,T)} \qquad (11)$$

It is instructive to compare in Figure 7 the theoretical prediction for the effective hydrodynamic radius with its experimental determination obtained by directly inverting the Stokes formula.

The free volume model correctly predicts that the hydrodynamic radius shows a maximum at the density at which the reduced mobility goes through the shallow local minimum. The density dependence of the hydrodynamic radius is expected to closely resemble the behavior of the system compressibility. Actually, when the compressibility is maximum, the attractive and repulsive forces in the system are balanced and the buildup of a complex ion-medium structure is favored. We recall that the position of the maximum is determined by the value of the α parameter that accounts for all the excess attractive interactions in the ion-medium system and that also determines the density of the maximum system compressibility. In a pure fluid the maximum compressibility occurs at the critical density whereas in the complex ion-medium system the strong ion-atom polarization interaction shifts the balance between attractive and repulsive forces to a much lower density.

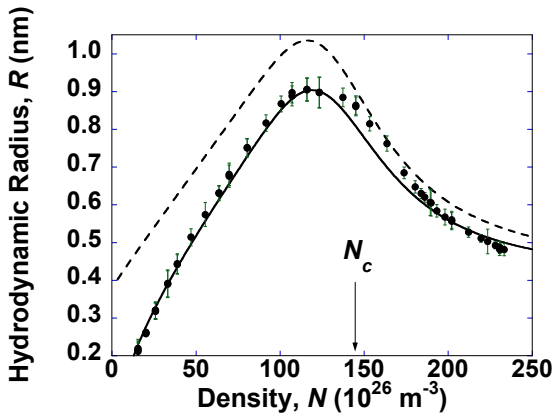

**Figure 7.** Density dependence of the effective hydrodynamic radius for $T$=45 K. Experiment (point) [15]. Lines: prediction of the free model with (solid line) or without (dashed line) the slip correction factor [22].

If the pure Stokes' formula is used, the free volume prediction for the effective hydrodynamic radius is roughly 20% too large than the experimental determination. However, if the slip factor correction is introduced, Equation (11) extremely well agrees with the data. We can thus conclude that a good description of the experimental data has to be based on both the use of the free volume model to compute the hydrodynamic ion radius and on the modification of the Stokes' formula by means of the Millikan-Cunningham slip factor correction.

With the values of the model parameters fixed by calibration on the 45 K isotherm data, we proceed to compute the reduced mobility for all other investigated isotherms. In Figure 8 we present the results for the three isotherms at 115 K, 190 K, and 270 K.

Although the density range investigated on these isotherms of higher temperature is not as broad as for the 45 K one, nonetheless, the agreement of the prediction of Equation (6) with the hydrodynamic radius given by Equation (7) is very good. The same good agreement has been obtained for all other isotherms in the range 64 K $\leq T \leq$ 330 K [23]. We have to stress the fact that the values of the free parameters have been set once for all by enforcing the agreement with the data on the lowest isotherm for which the investigated density range was the broadest.

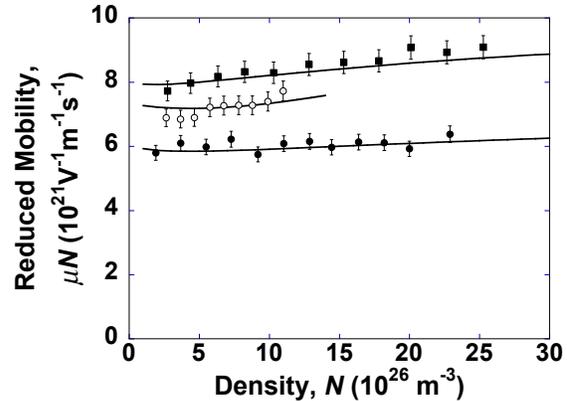

**Figure 8.** $\mu N$ vs $N$ for $T = 270$ K (open circles), 190 K (closed circles), and 115 K (squares).

In spite to the fact that the free volume model is based on an extension of the hydrodynamic Stokes' formula to lower densities, it is also able to correctly predict the temperature dependence of the zero-density limit of the reduced mobility. The solid line in Figure 4 is the prediction of the model obtained by assuming that the function $A(T)$ is the weakly temperature dependent function depicted in Figure 9.

## 4   CONCLUSIONS

The free volume model presented here is a thermodynamic approach to correctly computing the size of the complex which is formed when an anion is embedded in a host compliant medium (a gas and/or a liquid) as the consequence of the delicate balance between the long-range attractive polarization interaction between the ion and the host atoms and the short range repulsive exchange forces. The density behavior of the hydrodynamic radius shown in Figure 8 appears to depict the dynamics of negatively charged clusters whose structure consists of a cavity surrounded by a solvation shell. From the $O_2^-$ ion mobility measurements in a dense gas, Ne in the present case, a piece of information on the thickness of the solvation shell is gathered as the ion core does not change with density because the short range repulsive forces are very strong.

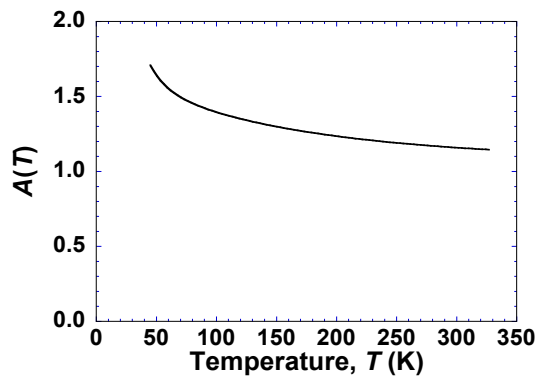

**Figure 9.** $A(T)$ vs $T$.

The free volume model is aimed at bridging the high-density hydrodynamic transport regime and the low-density kinetic regime. The deviations from the Stokes' law that grow increasingly larger when the density decreases are successfully treated by a very simple form of the Millikan-Cunningham slip correction factor.

The parameters in the free volume model need only to be fixed on a single isotherm provided that the available density range experimentally investigated is broad enough. Once this has been done the model can predict the mobility on all other isotherms.

It is worth noting that the parameters of the van der Waals-like description of the excess internal pressure due to the presence of the ions appear to be of universal validity. We believe that this universality might be a consequence of the Law of Corresponding States.

A drawback of the model is that it describes only the zero-field mobility but does not predict how the mobility should depend on the electric field. Actually, the experimental measurements do show a very weak electric field dependence, especially at low density and high temperature, which is probably due to a drift-induced deformation of the spherical symmetry of the free volume associated with the ions that is not described by the Kinetic Theory.

Finally, in order to test the validity of the present model in different systems, we are presently planning to analyze the mobility of $O_2^-$ ions in different dense gases.

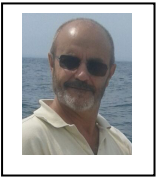 **Armando Francesco Borghesani** (born in Verona, Italy, 1956) graduated at the University of Padua, Italy, in 1979. In 1980-1981 he was appointed as junior Research Assistant at ST Microelectronics, where he was involved in the development of plasma etching processes for microelectronic devices. In 1981-1983, he served as Research Assistant at Eniricerche of the ENI group, where he studied the rheology of coal-based slurries. In 1983 he was appointed as Assistant Professor at the University of Padua and in 1992 he was promoted to Associate Professor. In his career, Prof. Borghesani was involved in research on the gas and liquid viscosity at critical point, on plasma etching, on the rheology of coal-oil/water mixtures, on electron and ion transport in dielectric fluids, on the infrared luminescence of electron-impact produced rare gas excimers and of particle beam excited rare-earth doped crystals.

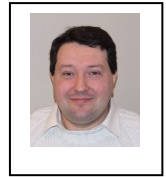 **Frédéric Aitken** (born in Grenoble, France, 1969) graduated in Mechanics from Grenoble Joseph Fourier University and Aeronautical Engineering School in Toulouse. In 1994-1995 he was appointed engineer at Institute Max Planck, where he was in charge of the study of cooling of resistive magnetic coils of Polyhelix type. He passed his Ph. D. degree in 1998 and joined the CNRS as a full researcher shortly after. 2010-2014, he was appointed as President of the French Physical Society (SFP-Alpes). His researches are focused on the investigation of fast energy injection in dielectric liquids for characterization of phase change induced by corona-like discharges. Making relations of his own with renowned scientists he became a specialist of non-equilibrium thermodynamic and transport properties in these fluids.